\documentclass{pp}
\newcommand{\beq}{\begin{equation}}
\newcommand{\eeq}{\end{equation}}
\newcommand{\beqa}{\begin{eqnarray}}
\newcommand{\eeqa}{\end{eqnarray}}
\newcommand{\ba}{\begin{array}}
\newcommand{\ea}{\end{array}}

\begin{document}

\title{Degenerate Quantum Gases \\
and Bose-Einstein Condensation} 
\author{Luca Salasnich} 
\address{
Istituto Nazionale per la Fisica della Materia, Unit\`a di Milano,\\ 
Dipartimento di Fisica, Universit\`a di Milano, \\
Via Celoria 16, 20133 Milano, Italy} 

\maketitle

\abstracts{After a brief historical 
introduction to Bose-Einstein condensation 
and Fermi degeneracy, we discuss 
theoretical results we have recentely obtained 
for trapped degenerate quantum gases 
by means of a thermal field theory approach. 
In particular, by using Gross-Pitaevskii and 
Bogoliubov-Popov equations, 
we consider thermodynamical properties 
of two Bosonic systems: a gas of Lithium atoms 
and a gas of Hydrogen atoms. 
Finally, we investigate finite-temperature density 
profiles of a dilute Fermi gas of Potassium atoms 
confined in a harmonic potential.} 

\section{Introduction}

Bose-Einstein condensation (BEC) is the macroscopic occupation 
of the lowest single-particle state of a system of bosons. 
The concept of Bose condensation dates back to 1924. In that year 
it appeared a paper of Bose [1] about a new 
derivation of photon statistic and Planck distribution 
and the same year Einstein [2] 
gave the theoretical description of BEC for a homogeneous system 
of identical particles. Einstein used the de Broglie 
matter wave for applying Bose calculation and developing 
a statistic also for particles. 
In 1938 London proposed the idea of a macroscopic wavefunction 
to describe the behavior of a Bose-condensed fluid [3]. 
His idea was that BEC should be involved in the phase 
transition of superfluid $^{4}$He. From this suggestion 
was derived the Tisza's two-fluid concept [4] and, 
consequently, the Landau's model [5]. 
Other contributions came from the Russian school 
(Bogoliubov, Beliaev, Popov). In particular 
the study of BEC and its elementary excitations  
in a weakly interacting dilute Bose gas [6]. 
\par 
From 1995 we have strong experimental 
evidence of BEC in clouds of confined alkali-metal 
atoms at ultra-low temperature (about $100$ nK): 
at JILA with $^{87}$Rb atoms [7], 
at MIT with $^{23}$Na atoms [8], and at Rice University 
with $^{7}$Li atoms [9]. In these experiments 
the condensed fraction can be more than $90\%$. 
More recently, another MIT group has obtained BEC 
(but small condensed fraction) with a gas of hydrogen atoms [10]. 
\par 
For dilute vapors one can separate off the 
effect of BEC and study the role of the interaction. 
The theoretical understanding of recent 
experiments is based on the Gross-Pitaevskii equation 
of a space-time dependent macroscopic wavefunction 
(order parameter) that describes the Bose condensate [11]. 
\par 
The experiments with alkali-metal atoms 
generally consist of three steps:
a laser cooling and confinement in an external potential 
(a magnetic or magneto-optical trap), 
an evaporative cooling and finally the analysis of 
the state of the system. 
It is important to observe that 
the Nobel Prize 1997 has been given for 
the development of methods to cool and 
trap atoms with laser light [12]. 
Nowadays more than twenty experimental groups have achieved BEC by using 
different geometries of the confining trap and atomic species. 
\par 
It is interesting to note that the paper of Einstein 
on BEC [2] preceded the concept of Fermi-Dirac statistics [14] 
as well as the division of all particles into two classes 
(Bosons and Fermions) depending on their net spin [15]. 
\par 
Very recently the Fermi quantum degeneracy has been 
obtained with dilute vapors of laser-cooled $^{40}$K 
atoms confined in a magnetic trap [15]. 
This experiment has renewed the interest on Fermi gases. 
In fact, with an effective attractive interatomic interaction,  
it should be possible to produce Cooper pairs of Fermions 
with Bosonic statistics and to investigate the mechanism 
of superfluidity with Fermions. 
\par 
In this paper we investigate 
dilute quantum gases at ultra-low temperature 
and confined by an external potential. 
In section 2 we review some basic concepts 
of thermal field theory for Bosons. 
In sections 3 and 4 we apply thermal field theory 
to study a Bose gas of Lithium and Hydrogen atoms, 
respectively. In section 5 we discuss thermal field 
theory for Fermions. Finally, in section 6 we analyze 
some properties a dilute Fermi gas of Potassium atoms. 

\section{Thermal Field Theory for Bosons} 

Thermal field theory investigates 
the many-body dynamics of finite-temperature systems, 
whose underlying structure is represented by quantum fields. 
This theory is the combination of statistical mechanics 
and quantum field theory [16]. 
Here, we apply the thermal field theory for 
non-relativistic Bosons in the same hyperfine state. 
\par
The Heisenberg equation of motion of the bosonic field 
operator ${\hat \psi}({\bf r},t)$, which describes 
a non-relativistic system of confined and interacting 
identical atoms in the same hyperfine state, is given by 
\beq 
i\hbar {\partial \over \partial t} {\hat \psi}({\bf r},t) =
\Big[ -{\hbar^2\over 2m} \nabla^2 
+ U({\bf r}) \Big] {\hat \psi} ({\bf r},t)
+ \int d^3{\bf r}' 
{\hat \psi}^+({\bf r}',t) V({\bf r},{\bf r}') 
{\hat \psi}({\bf r}',t){\hat \psi}({\bf r},t) \; , 
\eeq 
where $m$ is the mass of the atom, 
$U({\bf r})$ is the confining external potential and 
$V({\bf r},{\bf r}')$ is the interatomic potential. 
\par
The Bosonic field operator must satisfy the following 
equal-time commutation rules 
\beq 
[{\hat \psi}^+({\bf r}',t),{\hat \psi}({\bf r},t)]=
\delta^3({\bf r}'-{\bf r}) \; , 
\eeq
\beq
[{\hat \psi}({\bf r}',t),{\hat \psi}({\bf r},t)]= 
[{\hat \psi}^+({\bf r}',t),{\hat \psi}^+({\bf r},t)]=0 \; ,
\eeq
where $\delta^3({\bf r})$ is the 3D Dirac delta function 
and $[{\hat A},{\hat B}]={\hat A}{\hat B}-{\hat B}{\hat A}$. 
\par
When the density of the atomic cloud is such that 
the scattering length and the range of the interatomic interaction 
are less than the average interatomic distance, 
the true interatomic potential can be 
approximated by a local pseudo-potential 
\beq 
V({\bf r},{\bf r}')=g\delta^3({\bf r}-{\bf r}') \; , 
\eeq
where $g={4\pi \hbar^2 a/m}$ 
is the scattering amplitude of the spin triplet channel 
($a$ is the s-wave scattering length). 
In this way the Hamiltonian operator ${\hat H}$ of 
the system reads 
\beq 
{\hat H}= \int d^3{\bf r} \; {\hat \psi}^+({\bf r}) 
\Big[ -{\hbar^2\over 2m} \nabla^2 
+ U({\bf r}) \Big] {\hat \psi} ({\bf r})
+ {1\over 2}g{\hat \psi}^+({\bf r}){\hat \psi}^+({\bf r})
{\hat \psi}({\bf r}){\hat \psi}({\bf r}) \; . 
\eeq 
Because the Hamiltonian $\hat H$ is invariant 
under the global U(1) field transformation 
\beq 
{\hat \psi}({\bf r}) \to  e^{i \alpha} {\hat \psi}({\bf r}) \; , 
\eeq 
the Noether theorem gives us 
the conservation of the particle number 
\beq 
{\hat N}= \int d^3 {\bf r} \; 
{\hat \psi}^+({\bf r}){\hat \psi}({\bf r}) \; . 
\eeq 
\par
In equilibrium satistical mechanics, the grand canonical 
partition function $Z$ of a quantum system of Hamiltonian 
$\hat H$ and conserved number $\hat N$ of particles 
is written as 
\beq 
Z= Tr\left[ e^{-\left( {\hat H} -\mu {\hat N} \right)/kT} 
\right] \; , 
\eeq
where $T$ is the temperature of the thermal reservoir, $k$ 
is the Boltzmann constant, and $\mu$ is the chemical potential [16]. 
The thermal average of an observable $\hat A$ is given by 
\beq 
\langle {\hat A} \rangle = {1\over Z} 
Tr\left[{\hat A} e^{-\left( {\hat H} -\mu {\hat N} \right)/kT} 
\right] \; .
\eeq 
Note that the chemical potential $\mu$ fixes the average 
total number of particles. 
In the grand canonical ensemble the system 
is described by an effective Hamiltonian 
${\hat H}-\mu {\hat N}$. Therefore, 
the chemical potential $\mu$ can be introduced with 
the following shift 
\beq 
{\partial \over \partial t} 
\to 
{\partial \over \partial t}+ {i\over \hbar}\mu  
\eeq
in the equation of motion of the field 
operator $\hat \psi({\bf r},t)$. In this way the equation 
of motion of the field operator at finite temperature 
is given by 
\beq 
i\hbar {\partial \over \partial t} {\hat \psi}({\bf r},t) =
\Big[ -{\hbar^2\over 2m} \nabla^2 
+ U({\bf r}) - \mu \Big] {\hat \psi} ({\bf r},t)
+ g {\hat \psi}^+({\bf r},t)
{\hat \psi}({\bf r},t){\hat \psi}({\bf r},t) \; .  
\eeq 
\par
In a Bosonic system one can separate Bose-condensed 
particles from non-condensed ones by using 
of the following {\it Bogoliubov prescription} 
\beq 
{\hat \psi}({\bf r},t)=\Phi({\bf r}) {\hat I} 
+{\hat \phi}({\bf r},t) 
\eeq 
where ${\hat I}$ is the identity operator and 
\beq 
\Phi({\bf r}) =\langle {\hat \psi}({\bf r},t) \rangle 
\eeq 
is the order parameter (macroscopic wavefunction) of the condensate. 
The field ${\hat \phi}({\bf r},t)$ is the fluctuation operator, 
which describes the non-condensed fraction of Bosonic atoms [6]. 
It is important to observe that the Bogoliubov prescription 
breaks the U(1) symmetry of the system, 
namely the number operator $\hat N$ is not 
a conserved quantity if $\Phi({\bf r})\neq 0$. 
For a fixed chemical potential $\mu$, it exist a critical 
temperature $T_{BEC}$, called BEC critical temperature, below 
which the order parameter is no more identically zero [17]. 
In this case one speaks of {\it spontaneous symmetry breaking} 
because the ground-state does not possess 
the initial symmetry of the system [18]. 
This phenomenon is also known as 
{\it off diagonal long-range order} because, 
in presence of BEC, the two-body density operator 
$\langle {\hat \psi}({\bf r}'){\hat \psi}({\bf r}) \rangle$  
must satisfy the Penrose-Onsager criterion 
\beq 
\lim_{|{\bf r}'-{\bf r}|\to \infty} 
\langle {\hat \psi}({\bf r}'){\hat \psi}({\bf r}) \rangle 
= \Phi^*({\bf r}') \Phi({\bf r}) 
\; , 
\eeq 
where $\Phi({\bf r})$ is the eigenfunction of 
the largest eigenvalue of the two-body density operator [19]. 
\par
The Bogoliubov prescription for the field operator 
${\hat \psi}({\bf r},t)$ enables us to write 
the three-body thermal average in the following way 
\beq 
\langle 
{\hat \psi}^+({\bf r},t) 
{\hat \psi}({\bf r},t){\hat \psi}({\bf r},t)
\rangle = 
|\Phi({\bf r})|^2 \Phi({\bf r}) + 
2 \tilde{n}({\bf r}) \Phi({\bf r}) + 
\tilde{m}({\bf r}) \Phi^*({\bf r}) + 
\tilde{s}({\bf r})  \; ,
\eeq
where $\tilde{n}({\bf r})=\langle 
{\hat \phi}^+({\bf r},t){\hat \phi}({\bf r},t)\rangle$ 
is the density of non-condensed particles, 
while $\tilde{m}({\bf r})=\langle 
{\hat \phi}({\bf r},t){\hat \phi}({\bf r},t)\rangle$ 
and $\tilde{s}({\bf r})=\langle 
{\hat \phi}^+({\bf r},t){\hat \phi}({\bf r},t)
{\hat \phi}({\bf r},t)\rangle$ are anomalous densities. 
Now, we obtain an equation for $\Phi({\bf r})$ 
by taking the thermal average in the equation of motion 
for the field operator ${\hat \phi}({\bf r})$. 
In this way we find 
\beq 
\Big[ -{\hbar^2\over 2m} \nabla^2 
+ U({\bf r}) \Phi({\bf r}) + 
g |\Phi({\bf r})|^2 + 2 g \tilde{n}({\bf r}) \Big] 
\Phi({\bf r}) + g \tilde{m}({\bf r})\Phi^*({\bf r}) 
+ \tilde{s}({\bf r}) = \mu \Phi({\bf r}) \; , 
\eeq 
that is the exact equation of motion of the Bose-Einstein 
order parameter $\Phi({\bf r})$. 
This is not a closed equation due to the presence 
of the non-condensed density $\tilde{n}({\bf r})$ 
and of the anomalous densities $\tilde{m}({\bf r})$ and 
$\tilde{s}({\bf r})$. 
Obviously, if the particles of the system are all 
in the Bose condensate then the non-condensed density 
and the anomalous densities are zero. In this case 
the previous equation becomes 
\beq 
\Big[ -{\hbar^2\over 2m} \nabla^2 
+ U({\bf r}) \Phi({\bf r}) + g |\Phi({\bf r})|^2 \Big] 
= \mu \Phi({\bf r}) \; , 
\eeq 
that is the so-called {\it Gross-Pitaevskii equation} [11]. 
This equation can be used to describe 
a dilute Bose gas near zero temperature, 
when the non-condensate is very small compared with the 
condensate fraction. 
Another possible, less drastic, simplification, 
neglects only the anomalous densities.  
Then the equation of motion 
of the Bose-Einstein order parameter $\Phi({\bf r})$ becomes 
\beq 
\Big[ -{\hbar^2\over 2m} \nabla^2 
+ U({\bf r}) \Phi({\bf r}) + 
g |\Phi({\bf r})|^2 + 2 g \tilde{n}({\bf r}) \Big] 
\Phi({\bf r})  = \mu \Phi({\bf r}) \; . 
\eeq 
Also this equation, that we call {\it finite-temperature 
Gross-Pitaevskii equation}, is not closed. We must add an equation 
for the non-condensed density $\tilde{n}({\bf r})$ 
by studying the fluctuation operator ${\hat \phi}({\bf r},t)$. 
\par 
The exact equation of motion of the 
fluctuation operator ${\hat \phi}({\bf r},t)$ is obtained 
by subtracting Eq. (16) to Eq. (11) and 
using Eq. (12). Such equation is given by 
$$ 
i\hbar {\partial \over \partial t} {\hat \phi}({\bf r},t) =
\Big[ -{\hbar^2\over 2m} \nabla^2 
+ U({\bf r}) - \mu + 2 g |\Phi({\bf r})|^2 \Big] 
{\hat \phi}({\bf r},t) + 
$$
\beq 
+ g \Big[ 
\Phi({\bf r})^2 {\hat \phi}^+({\bf r},t)
+ \Phi^*({\bf r}) {\hat \phi}({\bf r},t)^2 
+ 2 \Phi({\bf r}) {\hat \phi}^+({\bf r},t)
{\hat \phi}({\bf r},t) + 
\eeq
$$ 
+{\hat \phi}^+({\bf r},t) 
\Phi({\bf r})^2 {\hat \phi}({\bf r},t)^2
- {\tilde m}({\bf r})\Phi({\bf r})
-{\tilde s}({\bf r}) \Phi({\bf r})
\Big] 
\; .  
$$  
To find the thermal excitations above the condensate 
one can simply linearize the previous equation. 
This can be done by means of the mean-field approximation:  
\beq 
{\hat \phi}^+({\bf r},t)
{\hat \phi}({\bf r},t)
\simeq 
\langle {\hat \phi}^+({\bf r},t)
{\hat \phi}({\bf r},t)
\rangle = \tilde{n}({\bf r}) \; ,  
\eeq
\beq
{\hat \phi}({\bf r},t)
{\hat \phi}({\bf r},t)
\simeq 
\langle {\hat \phi}({\bf r},t)
{\hat \phi}({\bf r},t)
\rangle = \tilde{m}({\bf r}) \; , 
\eeq
$$
{\hat \phi}^+({\bf r},t)
{\hat \phi}({\bf r},t)
{\hat \phi}({\bf r},t)
\simeq 
2 \langle {\hat \phi}^+({\bf r},t)
{\hat \phi}({\bf r},t) \rangle 
{\hat \phi}({\bf r},t)
+ \langle 
{\hat \phi}^+({\bf r},t) 
{\hat \phi}({\bf r},t) 
{\hat \phi}({\bf r},t) 
\rangle = 
$$
\beq
= 2 \tilde{n}({\bf r}) {\hat \phi}({\bf r},t) 
+ \tilde{m}({\bf r}) {\hat \phi}^+({\bf r},t) 
+ \tilde{s}({\bf r}) 
\eeq 
Within the mean-field approximation, 
the linearized equation of motion of the fluctuation operator reads 
$$
i\hbar {\partial \over \partial t} {\hat \phi}({\bf r},t) =
\Big[ -{\hbar^2\over 2m} \nabla^2 + U({\bf r}) - \mu 
+ 2 g \big(|\Phi({\bf r})|^2+\tilde{n}({\bf r})\big) \Big] 
{\hat \phi}({\bf r},t) + 
$$
\beq
+ g \big( \Phi({\bf r})^2 + 
\tilde{m}({\bf r}) \big) {\hat \phi}^+({\bf r},t) \; .  
\eeq 
This equation remains very complex and 
some simplification is usually performed. 
One possible approximation, called 
{\it Bogoliubov standard approximation}, 
neglects the non-condensate densities 
($\tilde{n}({\bf r})=\tilde{m}({\bf r})=0$). 
This approximation is particularly useful near zero temperature: 
the condensate and its elementary excitations are separately calculated. 
Another possible, less drastic, simplification is 
called {\it Popov approximation}. 
It neglects only the correlation function $\tilde{m}({\bf r})$ 
and it is able to give a more accurate description 
of the non-condensate fraction. 
The Popov approximation is expected to be reliable in the whole range 
of temperature except near $T_c$, where mean-field theories are 
known to fail. 
\par 
In any case, due to the presence of the self-adjoint operator 
${\hat \phi}^+({\bf r},t)$ in the linearized 
equation of motion (23), the fluctuation operator 
must be expanded by using the so-called Bogoliubov transformation
\beq 
{\hat \phi}({\bf r},t)= \sum_j \Big[ u_j({\bf r})
e^{-iE_jt/\hbar}{\hat a}_j +v_j^*({\bf r})
e^{iE_jt/\hbar}{\hat a}_j^+ \Big] \; , 
\eeq 
where ${\hat a}_j$ and ${\hat a}_j^+$ are bosonic operators and 
the complex functions $u_{j}({\bf r})$ and $v_{j}({\bf r})$ 
are the wavefunctions of the so-called 
quasi-particle and quasi-hole excitations of energy $E_j$. 
The functions $u_j({\bf r})$ and $v_j({\bf r})$ satisfy the
normalization condition 
\beq
\int d^3{\bf r} \; [u_j^*({\bf r})u_k({\bf r})
- v_j^*({\bf r})v_k({\bf r})] = \delta_{jk} \; . 
\eeq 
By inserting Eq. (24) into Eq. (23) with $m({\bf r})=0$, 
we get the {\it Bogoliubov-Popov equations}   
$$
\Big[ -{\hbar^2\over 2m} \nabla^2+ U({\bf r}) -\mu
+ 2 g n({\bf r}) \Big] 
u_j({\bf r})+g \Phi({\bf r})^2 
v_j({\bf r})= E_j u_j({\bf r}) \; ,
$$
\beq
\Big[ -{\hbar^2\over 2m} \nabla^2+ U({\bf r}) -\mu
+ 2 g n({\bf r}) \Big] 
v_j({\bf r})+g \Phi({\bf r})^2 
u_j({\bf r})= - E_j v_j({\bf r}) \; ,
\eeq
where $n({\bf r})=|\Phi({\bf r})|^2+{\tilde n}({\bf r})$ 
is the total density. 
\par 
Eq. (18) and Eq. (26) are supplemented by the relation
fixing the total average number of atoms in the system 
\beq
N=\int d^3 {\bf r} [n_0({\bf r})+\tilde{n}({\bf r})] \; ,
\eeq
where 
$$
n_0({\bf r})= |\Phi ({\bf r})|^2 \; , 
$$ 
\beq
{\tilde n}({\bf r})=\sum_j \Big( |u_j({\bf r})|^2+|v_j({\bf r})|^2
\Big ) \langle {\hat a}^+_j {\hat a}_j \rangle 
+|v_j({\bf r})|^2 \; , 
\eeq
with 
\beq
\langle {\hat a}^+_j {\hat a}_j \rangle = 
{1\over e^{E_j/k T}-1} 
\eeq
the Bose factor at temperature $T$. 
Note that also at zero temperature there is a 
non-condensed density (quantum depletion), 
given by $\sum_j |v_j({\bf r})|^2$. 
\par 
When $kT$ is much larger than the lowest elementary 
excitation, one can use the {\it Hartree approximation} 
neglecting the quasi-hole excitations ($v_j({\bf r})=0$). 
In this way, the Bogolibov-Popov equations reduce 
to the following Hartree equations  
$$ 
\Big[ -{\hbar^2\over 2m} \nabla^2+ U({\bf r}) -\mu
+ 2 g n({\bf r}) \Big] 
u_j({\bf r}) = E_j u_j({\bf r}) \; ,
$$
and the non-condensed density is simply 
\beq
{\tilde n}({\bf r})=\sum_j {|u_j({\bf r})|^2 \over e^{E_j/k T}-1} 
\; . 
\eeq 
In this case the non-condensed density is equivalent 
to the thermal density because the quantum depletion is zero. 
If there is a large number of particles 
in the thermal cloud, the Hartree thermal density 
can be calculated through the quasi-classical approximation. 
In the quasi-classical approximation one uses the classical 
single-particle energy $E({\bf p}, {\bf r})={\bf p}^2/(2m) 
+U({\bf r})+2gn({\bf r})$ instead of the quantum energy $E_j$. 
Thus the thermal particles behave as non-interacting Bosons 
moving in the self-consistent effective potential 
$U({\bf r})+2gn({\bf r})$. Then the previous formula becomes 
\beq 
{\tilde n}({\bf r})=\int 
{d^3{\bf p}\over (2\pi\hbar)^3} 
{1\over e^{(E({\bf p}, {\bf r})-\mu)/kT}-1} 
={1\over \lambda^3}
\; g_{3/2}\left( e^{-(U({\bf r})+2gn({\bf r})-\mu )/kT}\right) \; ,
\eeq 
where $\lambda=(2\pi \hbar^2 /m k_B T)^{1/2}$ is the thermal length 
and 
\beq 
g_{\alpha}(z)=\sum_{k=1}^{\infty} {z^k \over k^{\alpha}} \; .
\eeq 
In the effective potential, the term $2gn({\bf r})$ 
is the Hartree mean-field generated by 
interactions with other atoms. 

\section{Bose gas of Lithium atoms} 

In this section we study the thermodynamics of a 
Bose-Einstein condensate of $^7$Li atoms,  
which is particularly interesting because of the 
attractive interatomic interaction: only a maximum 
number of atoms can form a condensate; beyond that the 
system collapses. Up to now 
there are only few experimental data supporting BEC 
for $^7$Li vapors with a limited condensate number. 
Because of the small number of condensed atoms, the size of the 
Bose condensate is of the same order of the resolution 
of the optical images. So it is difficult to obtain precise 
quantitative informations, like the number of condensed atoms 
and the BEC transition temperature [9]. 
\par
We consider an isotropic harmonic trap 
\beq 
U({\bf r})={1\over 2}m \omega^2 (x^2+y^2+z^2) \; , 
\eeq
and use trap parameters of the experiment [9]: 
$\omega=878$ Hz for the frequency of the isotropic harmonic trap and 
$a=-27a_0$ for the scattering length ($a_0$ is the Bohr radius). 
We measure the radial coordinate 
$r$ in units of the characteristic oscillator length 
$a_h=\sqrt{\hbar /(m \omega)}$ and the energies 
in units of energy oscillator quantum $\hbar \omega$. 
\par 
We concentrate our attention on 
density profiles and condensate fraction by using both 
the Bogoliubov and Popov approximations. 
Actually, the Popov method and its quasi-classical approximation 
are a good starting point to estimate the contribution 
of the correlations in the gas for any 
finite value of the temperature.  
\par 
In the case of negative scattering length, 
a simple variational calculation with a Gaussian trial wavefunction  
shows that a harmonic trap supports a Bose condensate with 
a number of bosons smaller than a $N_c =0.67(a_h/|a|)$ [21]. 
In a homogeneous gas such critical number is zero. 

\begin{figure}[t]
\epsfxsize=25pc 
\epsfbox{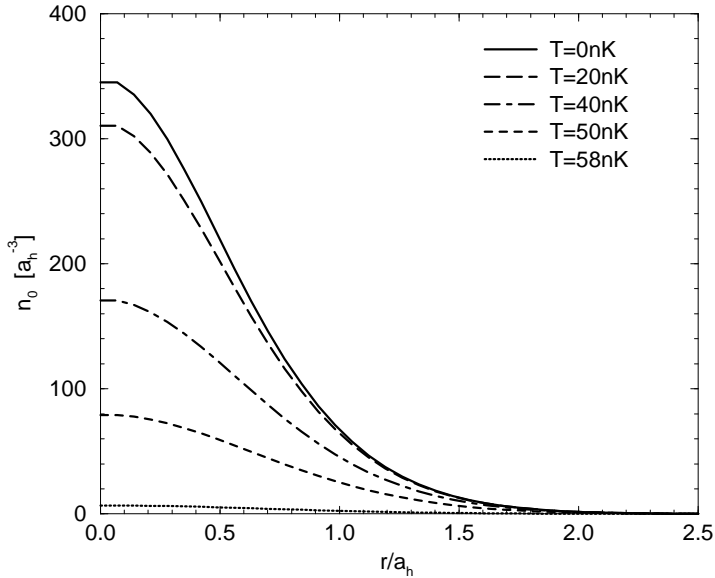} 
\caption{Popov calculations for $10^3$ $^7$Li atoms. 
Profile of the condensate density for temperature ranging 
from $0$ to $58$nK.}
\end{figure}

\par 
The system with $N=1000$ atoms 
is sufficiently far from the critical threshold $N_c = 1260$. 
We solve self-consistently the finite-temperature 
Gross-Pitaevskii equation (18) 
and the Bogoliubov-Popov equations (26) for some value of the temperature. 
In this way, we obtain informations about the density profiles 
of the condensate and non condensate fractions 
and the spectrum of elementary excitations [20]. 
\par
We plot in Fig. 1 the profile of the condensate fraction for some 
value of temperature from $0$ to $58$nK. Note that 
the temperature does not modify the shape of the condensate density and 
its radial range. This is not the case of the non condensed 
density profile, for which the increasing temperature produces a  
sharper maximum at a larger value of $r$.  

\begin{figure}[t]
\epsfxsize=25pc 
\epsfbox{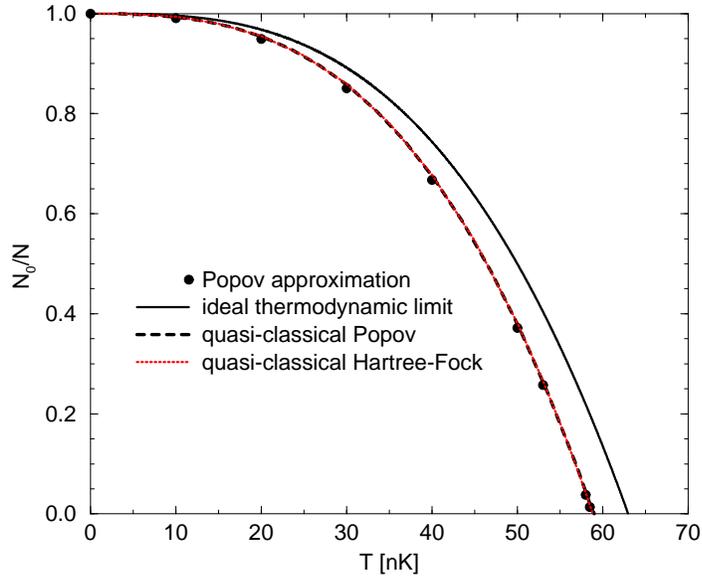} 
\caption{Condensate fraction {\it vs} temperature for 
$10^3$ $^7$Li atoms.}
\end{figure}

\par
In Fig. 2 we plot the condensate fraction $N_0/N$ versus the 
temperature $T$. In this figure we include also the 
quasi-classical approximation of the Popov and Hartree method 
and the curve of the ideal gas in the thermodynamic limit. 
We observe that the quasi-classical Popov 
approximation, for which the classical energy reads 
$E({\bf p},{\bf r})= \sqrt{ \big[{\bf p}^2/(2m) 
+U({\bf r})+2gn({\bf r})\big]^2 +g^2 n_0({\bf r}) }$ 
(see also [20]), gives accurate results 
for the condensate fraction at all temperatures here considered. 
\par
Now we analyze a total number of atoms near 
the critical threshold $N_c\simeq 1260$, at which there is the 
collapse of the condensate.  
We use the quasi-classical Popov approximation 
to evaluate the critical threshold $N_c$ as a function of 
the temperature. 

\begin{figure}[t]
\epsfxsize=25pc 
\epsfbox{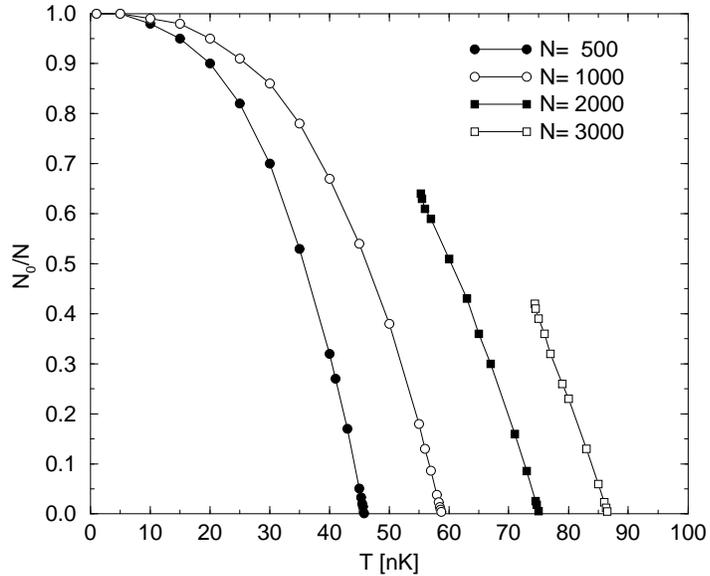} 
\caption{Condensate fraction {\it vs} temperature for 
different numbers of $^7$Li atoms. Results obtained with 
the quasi-classical Popov approximation. }
\end{figure}

\par
In Fig. 3 we plot the condensate fraction as a function of the 
temperature $T$ for different numbers of $^7$Li atoms. 
By increasing the temperature 
it is possible to put a larger number of atoms 
in the trap but the metastable condensate has a decreasing 
number of atoms. Also the critical number $N_0^c$ 
of condensed atoms decreases by increasing the temperature. 
These results, obtained with the quasi-classical Popov approximation, 
show that, when $\omega= 878$ Hz, for $N<N_c=1260$ 
the system is always metastable and it has a finite 
condensate fraction for $0\le T<T_{{}_{BEC}}$, 
where $T_{BEC}$ is the BEC transition temperature.  
For $N>N_c$ the system is metastable only for 
temperatures that exceed a critical value $T_c$ and it has 
a finite condensate fraction for $T_c<T<T_{{}_{BEC}}$ [20]. 
\par 
We have seen that when the condensate fraction is only few percents 
then $\tilde{n}({\bf 0})/n_0({\bf 0})$ becomes 
sufficiently large and the BEC cannot take place anymore. 
Thus, our calculations suggest that 
there should be a critical temperature $T^*$ 
beyond which the BEC transition 
is inhibited, independently of the number of atoms in the trap. 
Obviously, $T_c$ and $T_{{}_{BEC}}$ are functions of $N$. 
Such temperatures, and also the critical parameter $N_c$, 
are functions of the trap geometry, namely the frequency 
$\omega$, and of the scattering length $a$ [20]. 
\par 
Finally, we observe that our calculations could be useful 
to select experimental conditions for which a fraction 
of condensed $^7$Li atoms is detectable at thermal equilibrium. 

\section{Bose gas of Hydrogen atoms} 

BEC has been also achieved with atomic Hydrogen 
confined in a Ioffe-Pritchard trap [10]. That is an important 
result because Hydrogen properties, like interatomic 
potentials and spin relaxation rates, 
are well understood theoretically. 
The s-wave scattering length of the Hydrogen 
is very low ($a=0.0648$ nm) and, compared with other atomic species, 
the condensate density is high, even for small condensate fractions. 
Moreover, due to hydrogen's small mass, the BEC transition occurs 
at higher temperatures than previously observed. 
\par
We analyze the thermodynamical properties 
of the trapped Hydrogen gas by using the quasi-classical 
Hartree approximation. This approach is justified 
by the very large number of atoms (about $10^{10}$) in the 
trap and by the relatively high temperatures involved (order of $\mu$K). 
Our detailed theoretical study of the hydrogen 
thermodynamics can give useful informations for future experiments 
with a better optical resolution 
and a larger condensate fraction [22]. 
\par 
In the experiment reported in [10], the axially symmetric 
magnetic trap is modelled by the Ioffe-Pritchard potential 
\beq 
U(\rho , z) = \sqrt{(\alpha \rho)^2 + 
(\beta z^2+\gamma)^2} - \gamma  \; ,
\eeq 
where $\rho$ and $z$ are cylindrical coordinates and 
the parameters $\alpha$, $\beta$ and $\gamma$ 
can be calculated from the magnetic coil geometry. 
In particular, for small displacements, 
the radial oscillation frequency 
is $\omega_{\rho}=\alpha /\sqrt{m\gamma}=2\pi \times 3.90$ kHz, 
the axial frequency is $\omega_z=\sqrt{2\beta /m}=
2\pi \times 10.2$ Hz and $\gamma/k_B=35$ $\mu$K [10]. 
\par
It is important to observe 
in the experiments with atomic Hydrogen, the gas is dilute 
($n a^3<<1$) but also strongly interacting because 
$N a/a_h >> 1$, where 
$a_h=(\hbar /m \omega_h)^{1/2}$ with $\omega_h=
(\omega_{\rho}^2\omega_z)^{1/3}$. It follows that the kinetic 
term of the finite-temperature Gross-Pitaevskii 
equation (18) can be safely neglected 
and one gets the Thomas-Fermi condensate density 
\beq 
n_0({\bf r}) = {1\over g} 
\left[\mu -  U({\bf r}) - 2 g \tilde{n}({\bf r}) \right] \; ,
\eeq
in the region where $\mu > U({\bf r}) + 
2 g n_T({\bf r})$, and $n_0({\bf r})=0$ outside. 
In practice, we study the BEC thermodynamics by solving  
self-consistently the previous equation and Eq. (31). 

\begin{figure}[t]
\epsfxsize=25pc 
\epsfbox{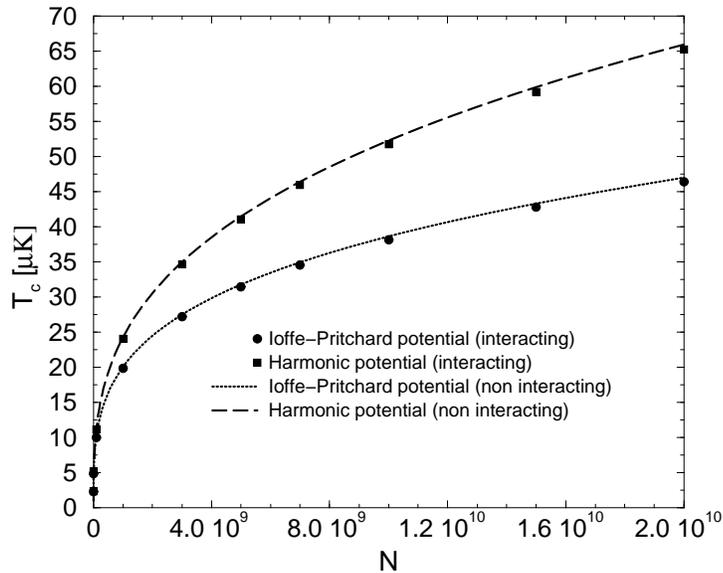} 
\caption{BEC transition temperature $T_c$ 
versus number $N$ of hydrogen atoms. Comparison between 
harmonic and Ioffe-Pritchard trapping potential.}
\end{figure}

In Fig. 4 we compare the BEC transition temperature $T_c^0$ 
of the Ioffe-Pritchard potential with the analytic formula 
$T_c^{0,H}=0.94 \hbar \omega_h N^{1/3}/k_B$, 
that is exact in the thermodynamic limit for 
the harmonic potential. Fig. 4 shows that, for a large number of atoms, 
$T_c^{0,H}$ exceeds $T_c^0$. In particular, 
for $N=2\cdot 10^{10}$ the relative difference is more 
than $41\%$. This strong effect is due to the larger 
density of states in the Ioffe-Pritchard trap. 

The role of the interatomic interaction 
on the transition temperature $T_c$ is very small. 
In Fig. 4 one observes that the repulsive interaction reduces $T_c$ 
both in harmonic and Ioffe-Pritchard traps. 
This effect is of the order of $0.1$ $\mu$K. 

\begin{figure}[t]
\epsfxsize=25pc 
\epsfbox{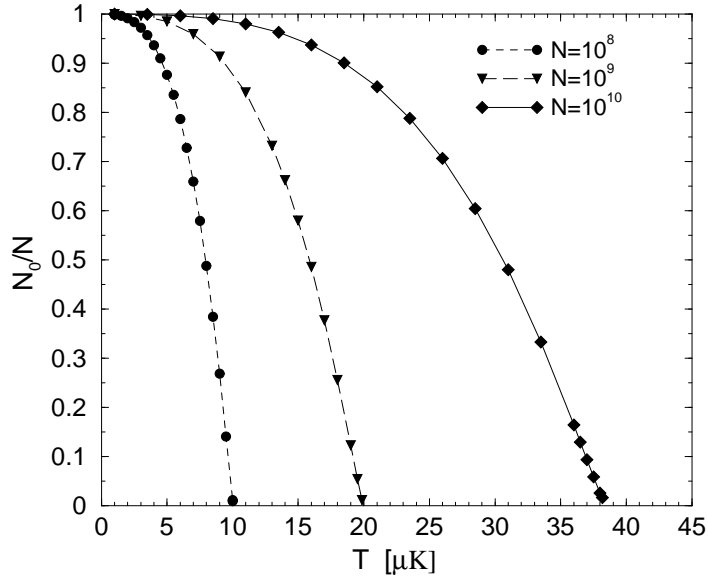} 
\caption{Condensate fraction $N_0/N$ as a function 
of temperature $T$. $N$ interacting hydrogen atoms 
in the Ioffe-Pritchard trap.} 
\end{figure} 

In Fig. 5 we plot the condensate fraction $N_0/N$ as a 
function of temperature 
for $N=10^8$, $10^9$ and $10^{10}$ atoms. 
As previously stated, such curves 
could be a useful guide for future experiments 
with a better optical resolution and 
a larger condensate fraction (further details 
can be found in [22]). 

\section{Thermal Field Theory for Fermions} 

The Heisenberg equation of motion of the fermionic field 
operator ${\hat \psi}_{\sigma}({\bf r},t)$, which describes 
 a non-relativistic system of confined and interacting 
identical atoms in the hyperfine state $\sigma$, is given by 
$$ 
i\hbar {\partial \over \partial t} {\hat \psi}_{\sigma}({\bf r},t) =
\Big[ -{\hbar^2\over 2m} \nabla^2 
+ U({\bf r}) \Big] {\hat \psi}_{\sigma}({\bf r},t) + 
$$
\beq 
+\sum_{\sigma'} \int d^3{\bf r}' 
{\hat \psi}_{\sigma'}^+({\bf r}',t) 
V_{\sigma'\sigma}({\bf r},{\bf r}') 
{\hat \psi}_{\sigma'}({\bf r}',t){\hat \psi}_{\sigma}({\bf r},t) \; , 
\eeq 
where $m$ is the mass of the atom, 
$U({\bf r})$ is the confining external potential and 
$V_{\sigma'\sigma}({\bf r},{\bf r}')$ is 
the interatomic potential [16]. 
\par 
The Bosonic field operator must satisfy the following 
equal-time anti-commutation rules 
\beq 
[{\hat \psi}_{\sigma'}^+({\bf r}',t),
{\hat \psi}_{\sigma}({\bf r},t)]_{+}= \delta_{\sigma'\sigma} 
\delta^3({\bf r}'-{\bf r}) \; , 
\eeq
\beq
[{\hat \psi}_{\sigma'}({\bf r}',t),{\hat \psi}_{\sigma'}
({\bf r},t)]_{+}= 
[{\hat \psi}_{\sigma'}^+({\bf r}',t),
{\hat \psi}_{\sigma}^+({\bf r},t)]_{+}=0 \; ,
\eeq
where $[{\hat A},{\hat B}]_{+}={\hat A}{\hat B}+{\hat B}{\hat A}$. 
\par
When the density of the atomic cloud is such that 
the scattering length and the range of the interatomic interaction 
are less than the average interatomic distance, 
the true interatomic potential can be 
approximated by a local pseudo-potential 
\beq 
V_{\sigma'\sigma}
({\bf r},{\bf r}')=g_{\sigma'\sigma}
\delta^3({\bf r}-{\bf r}') \; , 
\eeq 
where $g_{\sigma'\sigma}={4\pi \hbar^2 a_{\sigma'\sigma}/m}$ 
is the scattering amplitude and 
$a_{\sigma'\sigma}$ is the s-wave scattering length. 
The s-wave scattering between Fermions in the same hyperfine state 
is inhibited ($a_{\sigma\sigma}=0$) due to the Pauli principle. 
Nevertheless, the effect of interaction could be very effective  
for a Fermi vapor with two or more hyperfine states. 
Thus, in the grand canonical ensemble, the equation of motion 
of the field operator, which describes 
a dilute fermi gas with atoms in different 
hyperfine states, can be written as 
\beq 
i\hbar {\partial \over \partial t} {\hat \psi}_{\sigma}({\bf r},t) =
\Big[ -{\hbar^2\over 2m} \nabla^2 
+ U({\bf r}) -\mu \Big] {\hat \psi}_{\sigma}({\bf r},t)
+ \sum_{\sigma'} g_{\sigma'\sigma}
{\hat \psi}_{\sigma'}^+({\bf r},t) 
{\hat \psi}_{\sigma'}({\bf r},t){\hat \psi}_{\sigma}({\bf r},t) \; ,  
\eeq 
while in the case of a dilute Fermi gas of atoms in a unique hyperfine 
state, the previous equation is simply 
\beq 
i\hbar {\partial \over \partial t} {\hat \psi}({\bf r},t) =
\Big[ -{\hbar^2\over 2m} \nabla^2 
+ U({\bf r}) -\mu \Big] {\hat \psi}({\bf r},t) \; .  
\eeq 
This is the equation of motion of an ideal fermi gas 
in an external trapping potential. 
\par 
For an ideal Fermi gas in external potential 
the Fermionic field can be expandend 
in the following way 
\beq 
{\hat \psi}({\bf r},t)= \sum_{j} u_{j}({\bf r}) e^{-iE_jt/\hbar} 
{\hat a}_{j} \; , 
\eeq
where $u_{j}({\bf r})$ 
is the single-particle eigenfunction with eigenvalue 
$\epsilon_{j}$ and ${\hat a}_{j}$ is 
the lowering Fermi operator of the single-particle 
eigenstate $|j \rangle$. 
The thermal spatial density is given by 
\beq 
n({\bf r}) = 
\langle {\hat \psi}^+({\bf r},t){\hat \psi}({\bf r},t)\rangle 
= \sum_{j} |u_j({\bf r})|^2\langle{\hat a}^+_j {\hat a}_j \rangle 
\; , 
\eeq
with 
\beq 
\langle{\hat a}^+_j {\hat a}_j \rangle 
={1\over e^{E_j/k T}+1} 
\eeq 
the Fermi factor at temperature $T$. 
In the quasi-classical approximation one uses the classical 
single-particle energy $E({\bf p}, {\bf r})={\bf p}^2/(2m) 
+U({\bf r})$ instead of the quantum energy $E_j$.  
In this way the previous formula becomes 
\beq
n({\bf r})=\int 
{d^3{\bf p}\over (2\pi\hbar)^3} 
{1\over e^{(E({\bf p}, {\bf r})-\mu)/kT}+1} 
={1\over \lambda^3}
\; f_{3/2}\left( e^{-(U({\bf r})-\mu )/kT}\right) \; ,
\eeq
where $\lambda=(2\pi \hbar^2 /m k_B T)^{1/2}$ is the thermal length 
and 
\beq 
f_{n}(z) = {1\over \Gamma(n)} \int_0^{\infty} dx  
{x^{n-1}\over z^{-1}e^{x}+1} \; ,  
\eeq 
with $\Gamma(x)$ the factorial function. 
Note that for $|z|<1$ one has $f_{n}(z)=  
\sum_{i=1}^{\infty} (-1)^{i+1} z^i/i^n$.  
Moreover, by using $g_n(z)=-f_n(-z)$ instead of  
$f_n(z)$, one finds the spatial distribution  
of the ideal Bose gas in external potential. 
In the limit of zero temperature, with $\mu=E_F$ the Fermi energy, 
the quasi-classical spatial distribution 
is given by the Thomas-Fermi approximation 
\beq 
n({\bf r})=(2m)^{3/2}/(6\pi^2\hbar^3) 
\left(E_F- U({\bf r})\right)^{3/2}  
\Theta\left(E_F- U({\bf r})\right) \;, 
\eeq
where $\Theta$ is the Heaviside step function [17]. 

\section{Fermi gas of Potassium atoms} 

In 1999 the Fermi quantum degeneracy has been 
obtained with dilute vapors of laser-cooled $^{40}$K 
atoms confined in a magnetic trap [13]. 
In that experiment, to favor the evaporative cooling,  
a $^{40}$K Fermi vapor in two hyperfine states is used.  
When the system is below the Fermi temperature, 
one hyperfine component is removed and  
it remains a trapped quasi-ideal degenerate Fermi gas.  
The s-wave scattering between fermions in the same hyperfine state  
is inhibited due the Pauli principle.  
It follows that at low temperature the dilute Fermi gas,  
in a fixed hyperfine state, is practically ideal.  
Nevertheless, the effect of interaction could be very effective  
for a Fermi vapor with two or more hyperfine states. 
\par 
In the case of a harmonic external potential 
\beq 
U({\bf r}) = {1\over 2}m(\omega_1^2 x^2 + \omega_2^2 y^2  
+\omega_3^2 z^2) \;, 
\eeq 
one finds the Fermi density profile  
by using the Eq. (43) and 
the eigenfunctions $u_{n_1n_2n_3}({\bf r})$ 
of the harmonic oscillator  
\beq 
n({\bf r})= \sum_{n_1n_2n_3=0}^{\infty}  
{|u_{n_1n_2n_3}({\bf r})|^2  
\over e^{\beta\hbar(\omega_1(n_1+1/2)+\omega_2(n_2+1/2)  
+\omega_3(n_3+1/2)-\mu)} + 1} \; .  
\eeq  
Because the Fermi gas is ideal, one has 
$u_{n_1n_2n_3}({\bf r})=u_{n_1}(x) 
u_{n_2}(y)u_{n_3}(z)$, where $\phi_n(x)$ is the 
eigenfuction of a 1D harmonic oscillator with 
frequency $\omega$ and quantum number $n$. 
This eigenfunction can be found by means of 
the recursion relation 
\beq 
u_{n}(x) = {1\over \sqrt{n}} \Big[ 
\sqrt{2} a_h x u_{n-1}(x)- \sqrt{n-1} u_{n-2}(x) 
\Big] \; , 
\eeq
where $u_0(x)= a_h^{1/2}\pi^{1/4} e^{-a_h^2x^2/2}$ 
and $u_1(x)=\sqrt{2} a_h x u_0(x)$, 
with $a_h=\sqrt{\hbar/(m\omega)}$.  

\begin{figure}[t]
\epsfxsize=25pc 
\epsfbox{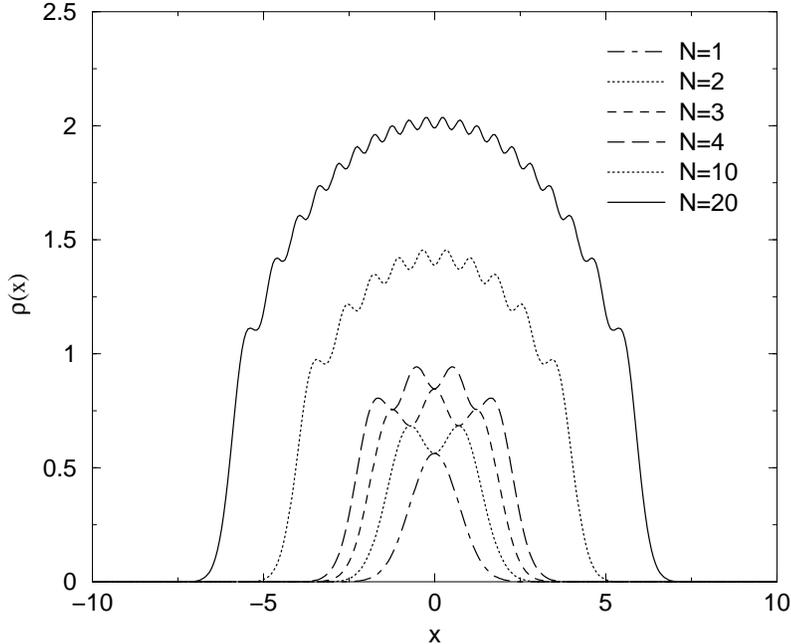} 
\caption{Density profiles for an ideal Fermi gas in a 1D harmonic trap.
Temperature: $kT/\hbar \omega =10^{-3}$; at this 
temperature the results coincide with zero-temperature ones. 
$N$ is the number of Fermions. 
Lengths in units $a_h=(\hbar/m\omega)^{1/2}$  
and densities in units $a_h^{-3}$.}
\end{figure}

\par 
In the quasi-1D case, namely a cigar-shaped 
gas where $\omega_1=\omega_2 >> \omega_3$, 
the shell effects are strongly enhanced. 
In Fig. 6 we plot the density profile of 
a 1D ideal Fermi gas in harmonic potential 
as a function of the number $N$ of particles 
at $kT/\hbar \omega=10^{-3}$. 
We have verified that at this temperature the 
density profiles coincide with the zero-temperature ones. 
The results are obtained by numerically evaluating 
expressions (49) and (50) in the 1D case. 
The local maxima, whose number grows with $N$, are clearly visible 
for a small number of particles. 

\begin{figure}[t]
\epsfxsize=25pc 
\epsfbox{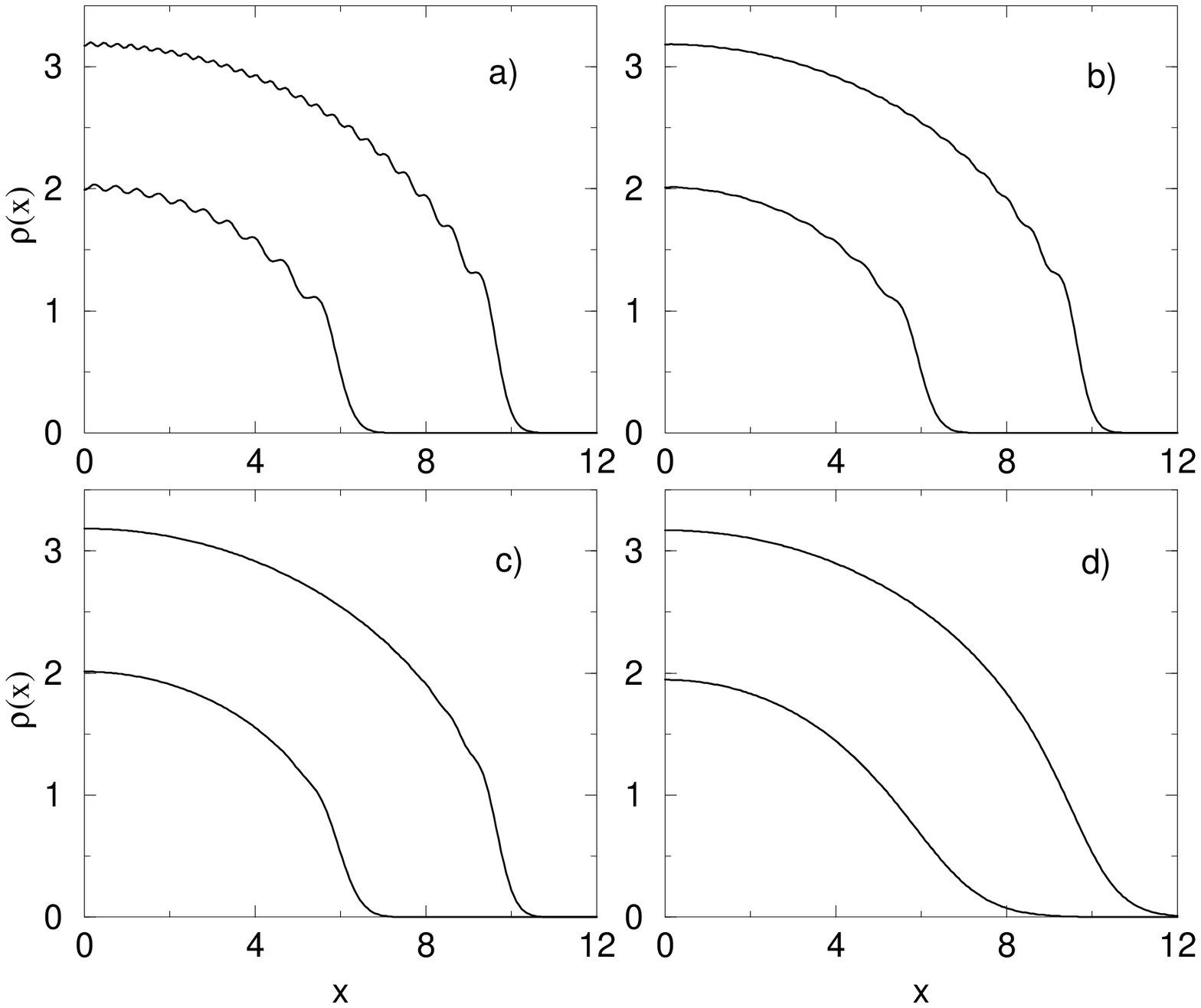} 
\caption{Density profiles for an ideal Fermi gas in a 1D harmonic trap 
at finite temperature, with $N=20$ and $50$. 
$N$ is the number of Fermions. a) $kT/\hbar \omega = 1/10$; 
b) $kT/\hbar \omega = 1/2$; c) $kT/\hbar \omega = 1$; 
d) $kT/\hbar \omega = 5$. Units as in Figure 6.}
\end{figure}

\par
In Fig. 7 we show the density profiles with $20$ and $50$ 
particles as a function of temperature. 
Remarkably, the local peaks are no more distinguishable 
for temperatures well below the Fermi temperature 
$T_F=N\hbar \omega/k$. Thus, to see spatial 
shell effect on the 1D density profile, the system 
should be at temperatures lower than $T_F$ 
by one or two orders of magnitude. 
\par 
The problem of a dilute Fermi vapor with $M$  
hyperfine states (components), 
can be studied by using the mean-field approximation  
and the quasi-classical formula.  
The spatial density profile $n_{\sigma}({\bf r})$ 
of the $\sigma$-th component 
in a $M$-component Fermi vapor can be written as  
\beq  
n_{\sigma}({\bf r})={1\over \lambda^3} 
f_{3/2}\left(e^{\left(\mu_{\sigma} -U({\bf r}) - 
\sum_{\sigma'=1}^{M-1} 
g_{\sigma'\sigma} n_{\sigma'}({\bf r}) \right)/kT}\right) \; ,  
\eeq 
where $\sigma=1,...,M$, $\mu_{\sigma}$ is the chemical potential 
of the $\sigma$-th component, 
and $g_{\sigma'\sigma}=4\pi\hbar^2a_{\sigma'\sigma}/m$, 
with $a_{\sigma'\sigma}$ the s-wave scattering length 
between $\sigma'$-th and $\sigma$-th  
component ($a_{\sigma\sigma}=0$). 
Thus, the effect of the other $M-1$ Fermi components on 
the $\sigma$-th component is the appearance 
of a mean-field effective potential. 
\par 
We numerically solve the set of equations (51) 
with a self-consistent iterative procedure. 
If the components of the Fermi vapor  
are non-interacting then they can occupy the same spatial region.  
Instead, if they are strongly interacting (repulsive interaction)  
there will be a phase-separation, i.e., the Fermi components  
will stay in different spatial regions. 
\par 
When two components have the same number of particles, the onset of  
phase-separation is also an example of spontaneous symmetry breaking.  
In fact, if the chemical potentials $\mu_i$ of the two components  
are equal, Eq. (51) always admits a symmetric solution  
$n_1({\bf r})=n_2({\bf r})$. However, for particle number $N$  
larger than a threshold $N_c$ the solution bifurcates and a pair of  
symmetry breaking solutions appears (see [23]). 

\begin{figure}[t]
\epsfxsize=25pc 
\epsfbox{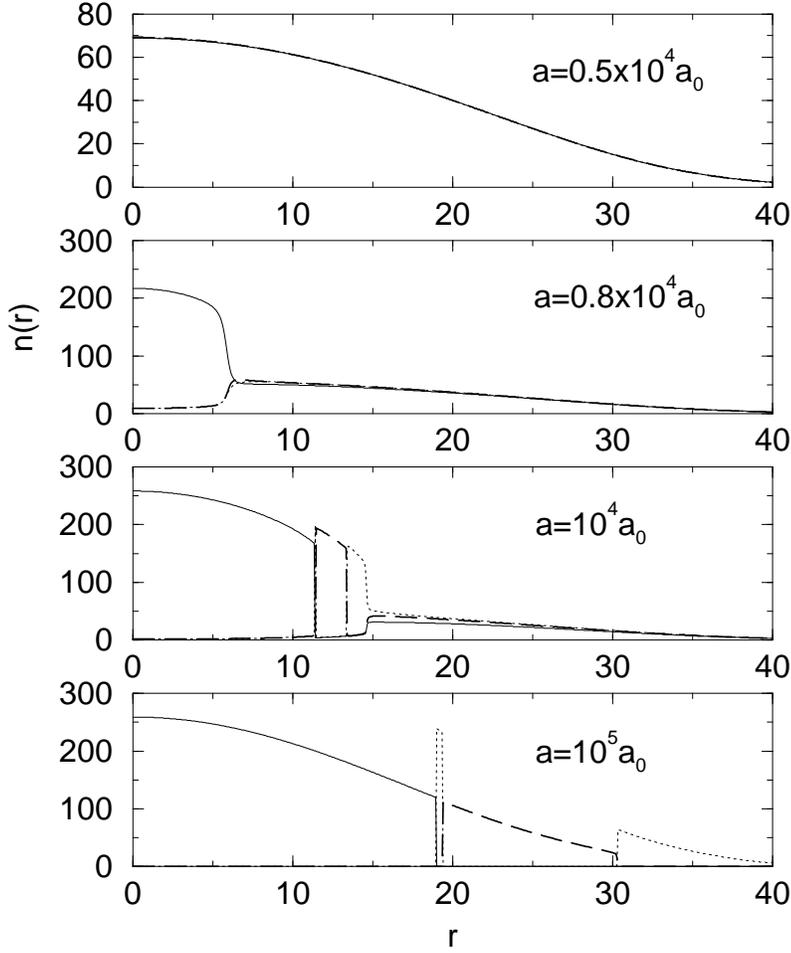} 
\caption{Density profiles of the  
$^{40}$K vapor with three components (solid, dotted and 
dashed lines) in a 3D isotropic harmonic trap ($\omega=450$ Hz).  
$N=0.5\cdot 10^7$ atoms for each component. 
Different values of the scattering length $a$ 
($a_0$ is the Bohr radius).  
Temperature $T=0.5 T_F$ ($T_F=1.07$ $\mu$K). 
Lengths in units $a_h=(\hbar/m\omega)^{1/2}$ 
and densities in units $a_h^{-3}$.}
\end{figure}

\par 
Phase-separation also appears in a Fermi vapor  
with three or more components. 
In Figure 8 we plot the density profiles of the $^{40}$K Fermi vapor  
with three components in a 3D isotropic 
harmonic potential. 
In this case we numerically solve Eq. (51) 
with $a_{12}=a_{13}=a_{23}=a$.  
The Figure shows that also for three components the  
phase-separation and the apparence 
of spatial shells are regulated by the scattering length.  
\par 
A Fermi vapor with more components has the  
same behavior. In particular, we have recently 
shown [23] that the critical density of fermions  
at the origin, which gives rise to the phase-separation,  
does not depend on the number of Fermi component  
and on the properties of the trap; moreover 
we have found, by using standard bifurcation theory,  
that such a density is given by $n_c({\bf 0})=\pi/(48 a^3)$,  
where $a$ is the s-wave scattering length. 

\section{Conclusions}

We have shown that non-relativistic 
quantum field theory is extremely useful 
to explain results of recent experiments with 
dilute alkali-metal atoms at ultra-low temperatures 
and to predict new quantum phenomena not 
yet experimentally observed. 
By studying these systems, 
sophisticated theoretical concepts, 
like order parameter, spontaneous symmetry breaking 
and off-diagonal long-range order, find a clear 
correspondence in measurable physical quantities. 
Finally, it is important to stress the role of 
the quasi-classical approximation, 
which is not only useful in computations but 
is also a giude for a deeper understanding 
of physical problems. 

\section*{Acknowledgements}
\par
The author thanks A. Ruffing and M. Robnik for their 
kind invitation to the International Conference 
'Bexbach Colloquium on Science'. 
The results presented here 
have been obtained in collaboration with 
B. Pozzi, A. Parola and L. Reatto. 

\section*{References}

\begin{description}

\item{\ [1]} Bose S N, 1924 Z. Phys. {\bf 26} 178 

\item{\ [2]} Einstein A, 1924 
Preussische Akademie der Wissenshaften 
{\bf 22} 261; 1925 {\bf 1} 3; 1925 {\bf 3} 18 

\item{\ [3]} London F, 1938 Nature {\bf 141} 643; 
1938 Phys. Rev. {\bf 54} 947; 1938 J. Phys. Chem. {\bf 43} 49 

\item{\ [4]} Tisza L, 1938 Nature {\bf 141} 913; 
1940 Journal de Physique et Radium {\bf 1} 164 350  

\item{\ [5]} Landau L D, 1941 J. Phys. U.S.S.R. {\bf 5} 71; 
1947 {\bf 11} 91 

\item{\ [6]} Bogoliubov N N, 1947 J. Phys. U.S.S.R. {\bf 11} 23; 
Beliaev S T, 1958 Sov. Phys. JETP {\bf 7} 299; 
Popov V N, 1987 {\it Functional Integrals and Collective 
Modes}, Cambridge University Press: Cambridge 

\item{\ [7]} Anderson M H, Ensher J R, Matthews M R, Wieman C E, 
and Cornell E A, 1995 Science {\bf 269} 189 

\item{\ [8]} Davis K B, Mewes M O, Andrews M R, van Druten N J,
Drufee D S, Kurn D M, and Ketterle W, 1995 
Phys. Rev. Lett. {\bf 75} 3969 

\item{\ [9]} Bradley C C, Sackett C A, Tollett J J, and Hulet R G, 
1995 Phys. Rev. Lett. {\bf 75} 1687 

\item{\ [10]} D.G. Fried D G, Killian T C, and Kleppner D, 
1998 Phys. Rev. Lett. {\bf 81} 3811 

\item{\ [11]} Gross E P, 1961 Nuovo Cimento {\bf 20} 454; 
1963 J. of Math. Phys. {\bf 2} 195; 
L.P. Pitaevskii, 1961 Sov. Phys. JETP {\bf 13} 451

\item{\ [12]} Chu S, Cohen-Tannouji C, and 
Phillips W D, 1997 Nobel Prize in Physics 
 
\item{\ [13]} Fermi E, 1926 Z. Physik {\bf 36} 902; 
Dirac P A M, 1926 Proc. Roy. Soc. A {\bf 112} 661 (1926)

\item{\ [14]} Fierz M, 1939 Helv. Phys. Acta {\bf 12} 3; 
Pauli W, 1940 Phys. Rev. {\bf 58} 716 

\item{\ [15]} DeMarco B and Jin D S, 1999 Science {\bf 285} 1703  

\item{\ [16]} Huang K, 1987 {\it Statistical Mechanics}, 
John Wiley: New York; Fetter A L and Walecka J D, 1971 
{\it Quantum Theory of Many-Particle Systems}, 
Mc Graw-Hill: Boston 

\item{\ [17]} Salasnich L, 2000 J. Math. Phys. {\bf 41} 8016 

\item{\ [18]} Goldstone J, 1961 Nuovo Cimento {\bf 19} 154; 
Anderson P W, 1966 Rev. Mod. Phys. {\bf 38} 298 

\item{\ [19]} Penrose O, 1951 Phil. Mag. {\bf 42} 1373; 
Penrose O and Onsager L, 1956 Phys. Rev. {\bf 104} 576  

\item{\ [20]} Pozzi B, Salasnich L, Parola A, and Reatto L, 
2000 J. Low Temp. Phys. {\bf 119} 57 

\item{\ [21]} Salasnich L, 1997 
Mod. Phys. Lett. B {\bf 11} 1249; 1998 {\bf 12} 649 

\item{\ [22]} Pozzi B, Salasnich L, Parola A, and Reatto L, 
2000 Eur. Phys. J. D {\bf 11} 367 

\item{\ [23]} Salasnich L, Pozzi B, Parola A, and 
Reatto L, 2000 J. Phys. B {\bf 33} 3943

\end{description}

\end{document}